\documentclass[a4paper,12pt]{article}
\usepackage{float,graphicx,subfigure}
\usepackage{amsmath,amssymb}
\usepackage[english]{babel}
\usepackage{natbib}
\usepackage{url}
\usepackage{stmaryrd}

\newcommand{\Fig}[1]{Fig. \ref{#1}}

\newcommand{\Eq}[1]{Eq.~(\ref{#1})}
\newcommand{\Tab}[1]{Tab. \ref{#1}}

\renewcommand{\vec}[1]{\mbox{\boldmath$#1$}}
\newcommand{\E}{\mathrm{e}}

\newcommand{\mng}[1]{\llbracket #1 \rrbracket}
\newcommand{\EV}{\mathbb{E}\,}

\bibpunct{(}{)}{,}{}{}{,}

\begin{document}

\title{\bf A neural network account to Kant’s philosophical aesthetics}
\author{
    Peter beim Graben\thanks{Bernstein Center for Computational Neuroscience, Berlin, Germany \\
    Email: peter.beimgraben@b-tu.de} \\
}
\date{\today}

\maketitle

\begin{abstract}
According to Kant’s (1724 -- 1804) philosophical aesthetics, laid down in his \emph{Critique of the Power of Judgement}
(1790), beauty is ``subjective purposefulness'', reflected by the ``harmony of the cognitive faculties'', which are ``understanding'' and ``imagination''. On the one hand, understanding refers to the mental capability to find regularities in sensory manifolds, while imagination refers to intuition, phantasy, and creativity of the mind, on the other hand. Inspired by the reinforcement learning theory of Schmidhuber, I present a neural network analogy for the harmony of the faculties in terms of generative adversarial networks (GAN) --- also often employed for artificial music composition --- by identifying the generator module with the faculty of imagination and the discriminator module with the faculty of understanding. According to the GAN algorithm, both modules are engaged in an adversarial game, thereby optimizing a particular objective function. In my reconstruction, the convergence of the GAN algorithm during the reception of art, either music or fine, entails the harmony of the faculties and thereby a neural network analogue of subjective purposefulness, i.e., beauty.
\end{abstract}

\vspace{-21cm}\hspace*{\fill}\begin{minipage}[b]{7cm}
\footnotesize
Wollt Ihr nach Regeln messen, \\
was nicht nach Eurer Regeln Lauf, \\
der eignen Spur vergessen, \\
sucht davon erst die Regel auf. \\
Richard Wagner, \emph{Die Meistersinger von N\"urnberg} (1868),
1st Act, 3rd Scene
\end{minipage}

\vspace{20cm}


\section{Introduction}
\label{sec:intro}

In October 2018, Christie's auction house in New York City had sold the painting \emph{Portrait of Edmond Belamy} by a formerly unknown artist with pseudonym ``GAN'' for \$432,500 \citep{Christies18}. The signature of the artist, drawn into the bottom right corner of the masterpiece, which reads
\begin{equation}\label{eq:ganopt}
  \min_G \max_D \EV_x [\log D(x)] + \EV_z [\log (1 - D(G(z)))] \:,
\end{equation}
instead of ``GAN'', indicates that the artist was an \emph{artificial intelligence (AI)}, called \emph{generative adversarial network} \citep{GoodfellowEA14}, a particular kind of a deep neural network \citep{CunBengioHinton15, Schmidhuber15}. The AI engineers behind ``GAN'' is the French collective \emph{Obvious}\footnote{
    \url{https://obvious-art.com/}
}
and the signature \eqref{eq:ganopt} essentially codifies the objective function of the network's training algorithm.

\begin{figure}[H]
\centering
 \subfigure[]{\includegraphics[width=0.4\linewidth]{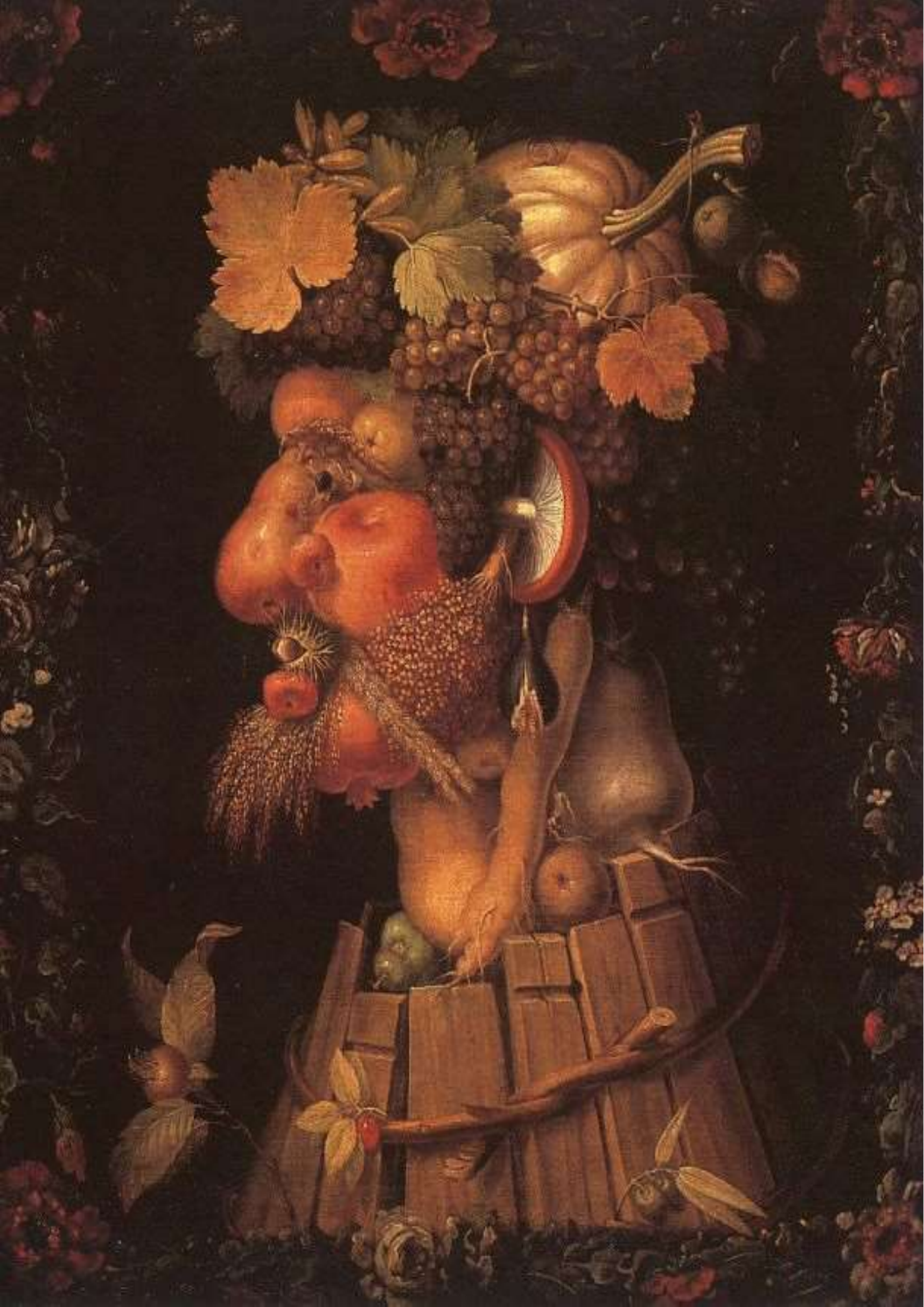}}
 \subfigure[]{\includegraphics[width=0.4\linewidth]{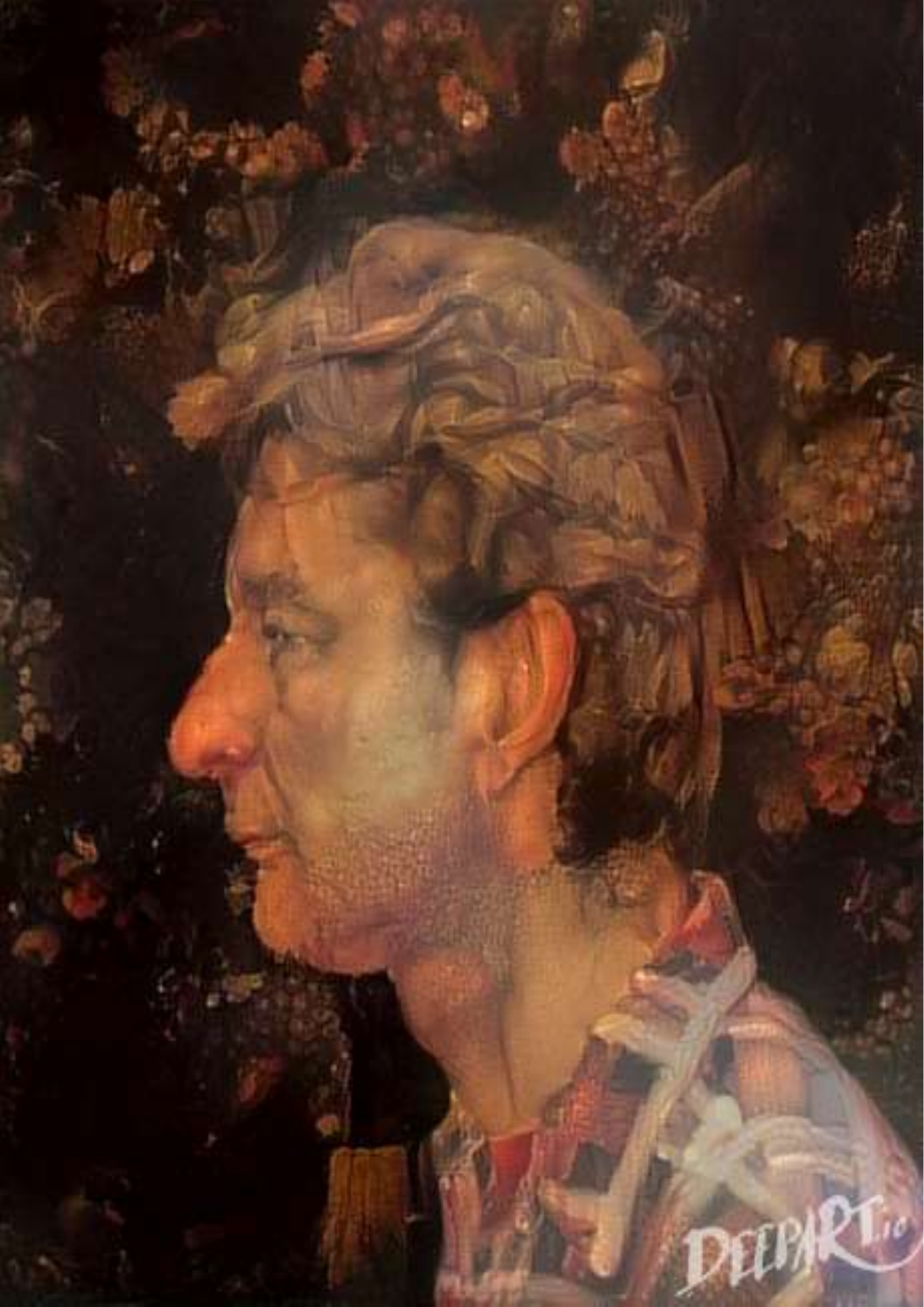}}
\caption{\emph{DeepArt} artistic playground. (a) Giuseppe Arcimboldo: \emph{Autumn}, 1573, Louvre, Paris. (b) \emph{DeepArt} converted portrait of the author, April, 25, 2016. (Color online) }\label{fig:deepart}
\end{figure}

Indeed, \emph{Obvious}' ``GAN'' was not the very first artistic AI application. One interesting precursor from the mid 2010-s was \emph{DeepArt},\footnote{
    \url{https://creativitywith.ai/deepartio/}
}
an initiative from the German University of T\"ubingen, for the image style transfer from specified examples to uploaded photographs  \citep{GatysEckerBethge16}. Figure \ref{fig:deepart} displays one of my own experiments with \emph{DeepArt} from April 2016. Here, \Fig{fig:deepart}(a) shows the painting \emph{Autumn} by mannerist artist Giuseppe Arcimboldo.\footnote{
    \url{https://en.wikipedia.org/wiki/Giuseppe_Arcimboldo#/media/File:Arcimboldo,_Giuseppe_~_Autumn,_1573,_oil_on_canvas,_Mus\%C3\%A9e_du_Louvre,_Paris.jpg}
}
The result from the style transfer to a portrait photograph of myself is shown in \Fig{fig:deepart}(b).

Obviously, the style transfer from Arcimboldo to the photograph had substantially failed. Instead of replacing my nose by a pea, my ear by a mushroom, or any other kind of palpable representational manipulation, the uploaded image was simply transformed by means of a low-pass filter, mostly blurring the details of my physiognomy. This is not really miraculous, since \emph{DeepArt}'s algorithm was a deep convolutional neural network (CNN) \citep{GatysEckerBethge16}, and thus a particular kind of a moving average filter \citep{RussellNorvig10}. More recently, numerous improved AI applications for fine art are available \citep{Murray19, CormackInverno12, VearPoltronieri22}, such as, e.g., \emph{Stable Diffusion} or \emph{Dall-E}.\footnote{
    \url{https://stability.ai/news/stable-diffusion-public-release}\\
    \url{https://openai.com/dall-e-3}
}

Deep neural networks in general and also generative adversarial networks in particular have found applications also in musical AI \citep{BriotHadjeresPachet20}. Specifically, \citet{Mogren16} used continuous recurrent neural networks (C-RNN) for artificial music composition,\footnote{
    For code and audio results of \citet{Mogren16}, see \url{https://mogren.one/publications/2016/c-rnn-gan/}
}
exploiting time-dynamical long short-term memory (LSTM) units \citep{HochreiterSchmidhuber97}, whereas the original proposal by \citet{GoodfellowEA14} essentially based on a perceptron architecture \citep{HertzKroghPalmer91}. Another approach by \citet{YangChouYang17} combined the GAN and CNN architectures for producing compositions in MIDI format, which was also used for network training.\footnote{
    For audio results of \citet{YangChouYang17}, see \url{https://soundcloud.com/vgtsv6jf5fwq/sets\%20}
}
As a last example for musical AI applications, I mention a network for melody generation from lyrics by \citet{YuSrivastavaCanales21}, also utilizing an LSTM architecture that is trained on MIDI examples.\footnote{
    See \url{https://www.youtube.com/watch?v=2PHcKhaLxAU} for an online demonstration of \citet{YuSrivastavaCanales21}
}

Now, the crucial question arises whether and to which extend, AI generated paintings or music pieces could be considered as \emph{real} art works \citep{Christies18}. In order to approach this question, one has to refer to empirical and philosophical aesthetics \citep{Gibbs22, MenninghausWagnerEA19, Tedesco24}.

For music, some recent ideas have been discussed within the framework of the ITPRA theory by \citet{Huron06}, who suggested ``five functionally distinct physiological systems: imagination, tension, prediction, reaction, and appraisal'' \citep[p.~7]{Huron06}. In this approach, the imagination module procures a ``simple act of daydreaming'' creating musical expectations (p.~8). Sometimes, expectations are delayed, e.g. by means of suspension, causing musical tension (pp.~307; 328). When an expected event actually occurs, the prediction module generates a rewarding emotion: ``listeners experience positive feelings whenever a future event is successfully predicted" (p.~ 239), whereas the opposite, a penalizing feeling of surprise takes place upon prediction failure within the phylogenetically ancient reaction module (pp.~13; 21). However, such ``inductive failures'' lead to an improvement of the prediction module by means of statistical learning (p.~217). Finally, the evolutionary most recent appraisal module may turn a bad reaction response into a consciously appreciated aesthetic emotion (p.~14), through ``contrastive valence'' (p.~239).

Huron's theory has been challenged and also partially confirmed in a recent experiment by \citet{CheungHarrisonEA19} (cf. the review of \citet{Huron19} and the thorough discussion by \citet{Blutner24b}). These researchers manipulated the predictability of chord cadences in either statistically certain or uncertain contexts. In agreement with Huron's ITPRA theory, aesthetic appreciation was high for predictable closings in rather uncertain contexts. However, \citet{CheungHarrisonEA19} also reported high pleasure ratings for surprising events in relatively certain contexts, which has to be attributed to the conscious appreciation response in Huron's theory. This interesting finding will be further addressed in the discussion.

The ITPRA theory by \citet{Huron06} was essentially inspired by the influential work of \citet{Meyer56}. For Meyer, ``aesthetic beliefs'' (p.~73) depend on ``embodied meaning'':
\begin{quote}
From this point of view what a musical stimulus or a series of stimuli indicate and point to are not extramusical concepts and objects but other musical events which are about to happen. That is, one musical event (be it a tone, a phrase, or a whole section) has meaning because it points to and makes us expect another musical event. [\dots] Embodied musical meaning is, in short, a product of expectation. If, on the basis of past experience, a present stimulus leads us to expect a more or less definite consequent musical event, then that stimulus has meaning. \citep[p.~35]{Meyer56}
\end{quote}

This view had been prepared by the musical aesthetics of \citet{Hanslick1891}:
\begin{quote}
The most important factor in the mental process which accompanies the act of listening to music, and which converts it into a source of pleasure, is frequently overlooked. We here refer to the intellectual satisfaction which the listener derives from continually following and anticipating the composer's intentions --- now, to see his expectations fulfilled, and now, to find himself agreeably mistaken. It is a matter of course that this intellectual flux and reflux, this perpetual giving and receiving takes place unconsciously, and with the rapidity of lightning-flashes. Only \emph{that} music can yield truly aesthetic enjoyment which prompts and rewards the act of thus closely following the composer's thoughts, and which with perfect justice may be called \emph{a pondering of the imagination}. Indeed, without mental activity no aesthetic enjoyment is possible. \citep[p.~135f]{Hanslick1891}
\end{quote}

Finally, I mention the book of \citet{Michaelis1795} which lays the ground for my present exposition. He wrote (in my own translation):
\begin{quote}
Perhaps, musical art could be declared as the art of stirring emotions, of vivifying and engaging the phantasy, and of tuning the mind towards ideas of the beautiful and the sublime through the diversified combination of tones; or briefly: as the art to immediately stimulate aesthetic emotions and aesthetic ideas by the conjunction of tones. \citep[pp.~54]{Michaelis1795}
\end{quote}

All those quotations together indicate how music reception (and the reception of art in general) could be regarded as a dynamical process: At one moment of time, a recipient entertains particular ``aesthetic beliefs'' \citep[p.~73]{Meyer56}, which are mental states comprising all relevant expectations, propensities, or schemata about the world, or shortly, \emph{belief states} in terms of \emph{dynamic semantics} \citep{Gardenfors88, Graben06, Graben14a} and partially observable decision processes in AI research \citep{RussellNorvig10}. Then, a musical episode (or a particular view upon a painting, a building, a dance, or a sculpture) is perceived that induces a transition from an ``antecedent'' state to a ``consequent'' state \citep[p.~26]{Meyer56}. Hence, the \emph{meaning} of the episode (or view) is represented by an \emph{operator} (like in quantum physics) on the space of belief states of a cognitive agent \citep{Graben06, Graben14a}.

Yet, the crucial keyword of the quotation from \citet[pp.~54]{Michaelis1795} is ``aesthetic ideas'' which directly refers to the context of my study, namely Kant's philosophical aesthetics, which is elucidated in the next section. More specifically, my present account does not aim at a literal, hermeneutic interpretation of Kant's aesthetic theory, but rather at a reconstruction of aesthetic reception in terms of contemporary scientific insights. In a first step, I use model-theoretic semantics in order to illustrate some of the main concepts of Kant's theory \citep{AchouriotiLambalgen11}. In a second step, I demonstrate how Kant's ideas lead straightforwardly to the intended reconstruction by means of artificial intelligence and neural network theory \citep{KimSchonecker22}. In particular, I show that aesthetic reception could be described in terms of generative adversarial network theory (GANT) \citep{GoodfellowEA14, Schmidhuber10a, Schmidhuber12}.\footnote{
    Compare also Schmidhuber's web blog \url{https://people.idsia.ch/~juergen/artificial-curiosity-since-1990.html}, containing a lot of technical reports on that issue.
}
By analogy, Kant's famous harmonious free play of the cognitive faculties \citep{Ginsborg97, Guyer06b} becomes isomorphic to the adversarial training regime of a GAN. The main focus of my study is art reception; the creation of art and the fundamental question whether AI might be considered as \emph{real} artists will be addressed in the concluding discussion.


\section{Kant's Analytic of the Beautiful}
\label{sec:kant}

The foundation of Kant's critical philosophy, presented in the three volumes \emph{Critique of Pure Reason} \citep{Kant99}, \emph{Critique of Practical Reason} \citep{Kant00}, and \emph{Critique of the Power of Judgement} \citep{Kant14}\footnote{
    Note that the \emph{Critique of the Power of Judgement} in the Cambridge Edition was not available for the present study. Therefore, I quote this book not, as usual, in concordance with the German \emph{Akademie Ausgabe}, but rather with the pagination of the translation by J. H. Bernard, entitled \emph{Kant's Critique of Judgement}, together with the (invariant) enumeration of paragraphs therein \citep{Kant14}. Further note that some of Kant's key concepts, such as ``Understanding'', ``Reason'', or ``Imagination'' are written in capital letters in this edition. Specifically, a distinction between ``Judgement'' [\emph{Urteilskraft}] (i.e. ``power of judgment'') and ``judgement'' [\emph{Urteil}] is made. For further comparison, consult the Kant glossary \citep[pp.~xlvii]{Kant14}. All other citations refer to the \emph{Akademie Ausgabe}, herein.
}
relies upon a tripartite ``division of the higher faculties of cognition'', which are ``Understanding'', ``Judgement'', and ``Reason'' \citep[B169]{Kant99}. ``Understanding'' denotes the ``faculty of rules'' \citep[B171]{Kant99}, or, in other words, the ``faculty of concepts'' \citep[§29, p.~131]{Kant14}. ``Judgement'', i.e. the ``power of judgment is the faculty of subsuming under rules'' \citep[B171]{Kant99}, and ``Reason'' refers to the ``faculty of Ideas'' \citep[§29, p.~137]{Kant14}, or, equivalently to the faculty ``for the derivation from principles'' \citep[p.~201]{Kant00}, i.e., the capability of logical deduction and consistence \citep{Stangneth19}.

In the \emph{Critique of the Power of Judgement}, Kant further differentiates between ``determinant'' Judgement \citep[§IV, p.~ 17]{Kant14}, and ``reflective'' Judgement (cf. also \citep[B171]{Kant99}):

\begin{quote}
Judgement in general is the faculty of thinking the particular as contained under the Universal. If the universal (the rule, the principle, the law) be given, the Judgement which subsumes the particular under it [\dots] is determinant. But if only the particular be given for which the universal has to be found, the Judgement is merely reflective. \citep[§IV, p.~ 17]{Kant14}
\end{quote}

While the determinant Judgement is the power of \emph{subsumption}, the reflective Judgement can be regarded as the capability of \emph{unification} \citep{RussellNorvig10}, driving the advancement of science as well as artistic enculturation of humanity \citep{Mizraji23}. Both aspects of mental life are tightly intermingled to emotionality:

\begin{quote}
For all faculties or capacities of the soul can be reduced to three, which cannot be any further derived from one common ground: the faculty of knowledge, the feeling of pleasure and pain, and the faculty of desire. \citep[§III, p.~15]{Kant14}
\end{quote}

Specifically, Kant drew here an analogy between Understanding as the ``faculty of knowledge'', reinforcing Judgement, and finally Reason as the (moral) ``faculty of desire''. Because Judgement is regulated by the transcendental idea of purposiveness \citep[§78, p.~331]{Kant14},

\begin{quote}
[\dots] the object is only called purposive, when its representation is immediately combined with the feeling of pleasure; and this very representation is an aesthetical representation of purposiveness. \citep[§VII, p.~31]{Kant14}
\end{quote}

An important prerequisite for cognition is the power of determinant judgement, which Kant models as a ``threefold synthesis'' of \emph{apprehension}, \emph{reproduction}, and \emph{recognition} \citep[A97]{Kant99} (cf. also \citet{AchouriotiLambalgen11}). More specifically, he argued:

\begin{quote}
Every empirical concept requires three acts of the spontaneous faculty of cognition: 1. The \emph{apprehension} [\emph{Auffassung}] (\emph{apprehensio}) of the manifold of intuition 2. the \emph{comprehension}  [\emph{Zusammenfassung}] i.e. the synthetic unity of the consciousness of this manifold in the concept of an object  (\emph{apperceptio comprehensiva}) 3. the \emph{exhibition} [\emph{Darstellung}] (\emph{exhibitio}) in intuition of the object corresponding to this concept. For the first act imagination is required, for the second understanding, and for the third judgement. \citep[p.~220]{Kant00}
\end{quote}

Productive imagination (\citet[B181]{Kant99}, \citet[§49, p.~198]{Kant14}) is likewise regarded as ``the faculty of presentation'' \citep[§17, p.~85]{Kant14}, or ``as faculty of intuition'' \citep[§39, p.~168]{Kant14}, sometimes governed by the ``laws of Association'' \citep[§29, p.~136]{Kant14}.

A corresponding combination of imagination, understanding and reflective judgement is also crucial for aesthetic experiences where ``the Imagination is here creative'', which ``gets to fiction [\dots] in the peculiar fancies with which the mind entertains itself'' \citep[§22, p.~100]{Kant14}. Thus, one may interpret Imagination as the productive and creative faculty of the human mind.

Now, the central idea of Kant's analytic of the beautiful can be captured by the following passage:

\begin{quote}
The consciousness of the mere formal purposiveness in the play of the subject's cognitive powers, in a representation through which an object is given, is the pleasure itself; because it contains a determining ground of the activity of the subject in respect of the excitement of its cognitive powers, and therefore an inner causality (which is purposive) in respect of cognition in general without however being limited to any definite cognition; and consequently contains a mere form of the subjective purposiveness of a representation in an aesthetical judgement. \citep[§12, p.~71]{Kant14}.
\end{quote}

Hence, a person judges a given object \emph{beautiful}, when its perception excites a ``free play'' of her ``cognitive powers'', Imagination and Understanding \citep[§9, p.~64]{Kant14}. When this interaction appears as being ``harmonious'' \citep[§39, p.~168]{Kant14}, it is felt pleasurable \citep[§15, p.~80]{Kant14}, and eventually the object is regarded as being subjectively purposive \citep[§11, p.~69]{Kant14}, meaning that it has no other end than eliciting pleasure through the subjectively experienced harmony of the free play between Imagination and Understanding.

Kant also used another paradoxical phrasing:\footnote{
    Similarly, citing \citet[p.~219]{Wagner84}: ``Only through phantasy, understanding is able to consort with emotion.'' [\emph{``Nur durch die Phantasie vermag der Verstand mit dem Gefühle zu verkehren.''} My translation]
}

\begin{quote}
Now if in the judgement of taste the Imagination must be considered in its freedom, it is in the first place not regarded as
reproductive, as it is subject to the laws of association, but as productive and spontaneous [\dots] The Understanding alone gives the law [\dots] Hence it is a conformity to law without a law; and a subjective agreement of the Imagination and Understanding, without such an objective agreement as there is when the representation is referred to a definite concept of an object, can subsist along with the free conformity to law of the Understanding (which is also called purposiveness without purpose) and with the peculiar feature of a judgement of taste. \citep[§22, p.~96]{Kant14}
\end{quote}

Yet, taste as the sense for the beautiful does not appear absolutely useless in Kant's philosophy, because it ``brings with it a feeling of the furtherance of life'' \citep[§23, p.~102]{Kant14}, and therefore ``quickens the cognitive faculties'' \citep[§49, p.~201]{Kant14}, eventually promoting ``the feeling of health'' \citep[§54, p.~221]{Kant14}.

Consequently, Kant defined ``aesthetical Ideas'' through:

\begin{quote}
Spirit, in an aesthetical sense, is the name given to the animating principle of the mind. But that whereby this principle
animates the soul, the material which it applies to that [purpose], is that which puts the mental powers purposively into swing, i.e. into such a play as maintains itself and strengthens the [mental] powers in their exercise. Now I maintain that this principle is no other than the faculty of presenting aesthetical Ideas. And by an aesthetical Idea I understand that representation of the Imagination which occasions much thought, without, however, any definite thought, i.e. any concept, being capable of being adequate to it; it consequently cannot be completely compassed and made intelligible by language. \citep[§49, p.~197]{Kant14}.
\end{quote}

Which can finally be related to music:

\begin{quote}
On the other hand, music and that which excites laughter are two different kinds of play with aesthetical Ideas, or with representations of the Understanding through which ultimately nothing is thought; and yet they can give lively gratification merely by their changes. \citep[§54, p.~222]{Kant14}.
\end{quote}

In this sense, Kant had 	paved the way for the music aesthetics of \citet{Michaelis1795}, \citet{Hanslick1891}, \citet{Meyer56}, and \citet{Huron06}, discussed above in terms of dynamic semantics \citep{Gardenfors88, Graben06, Graben14a}.


\subsection{Model Theory on Kant}
\label{sec:modelkant}

Kant's metaphors and paradoxical formulations as summarized above have presented hermeneutic challenges to his interprets (e.g. \citet{Ginsborg97, Ginsborg03, Guyer06b}). \Citeauthor{Ginsborg03}, e.g., gave the following interpretation

\begin{quote}
[\dots] first the feeling of pleasure, which judges the subjective purposiveness of the object or its representation for our cognitive faculties; second, the judgment which ascribes universal validity to the pleasure; and third, the still higher-order judgment which claims that the previously mentioned judgment is pure and hence, itself, universally valid [\dots]  \citet[p.~169]{Ginsborg03}
\end{quote}
for some passages in \citet[§9, p.~65; §35, p.~161]{Kant14}, while \citet{Guyer06b} distinguished between \emph{precognitive}, \emph{multicognitive}, and --- his own suggestion --- \emph{metacognitive} \citep[p.~182]{Guyer06b} approaches for the interpretation of the ``harmony of the cognitive faculties'' \citep[§9, p.~65]{Kant14}.

In order to avoid such interpretational problems and for further illustrating Kant's essential ideas of aesthetic philosophy, I propose a model-theoretic account in the present subsection \citep{AchouriotiLambalgen11}. In model-theoretic semantics, judgements are simply propositions. Thus, it suggests itself to consider propositional logic. Moreover, for understanding is the ``faculty of concepts'' \citep[§29, p.~131]{Kant14}, and concepts can be identified with predicates of first-order predicate logic, one has to consider predicate logic as well. The syntax of predicate logic is prescribed as a term algebra over the disjoint symbol sets of variables, constants, predicates, and logical operators, comprising the quantors and the connectives of propositional logic. Then, its model-theoretic semantics is basically given by an interpretation function, mapping constants onto elements of a suitably chosen discourse domain $M$ of individual entities and mapping predicates onto relations over the direct products of $M$. In this way, the model-theoretic \emph{meaning} of a unary predicate is given as a subset of the discourse domain $M$, or, in other words, predicates are interpreted extensionally \citep{RussellNorvig10}.

\begin{figure}[H]
\centering
 \subfigure[]{\includegraphics[width=0.5\linewidth]{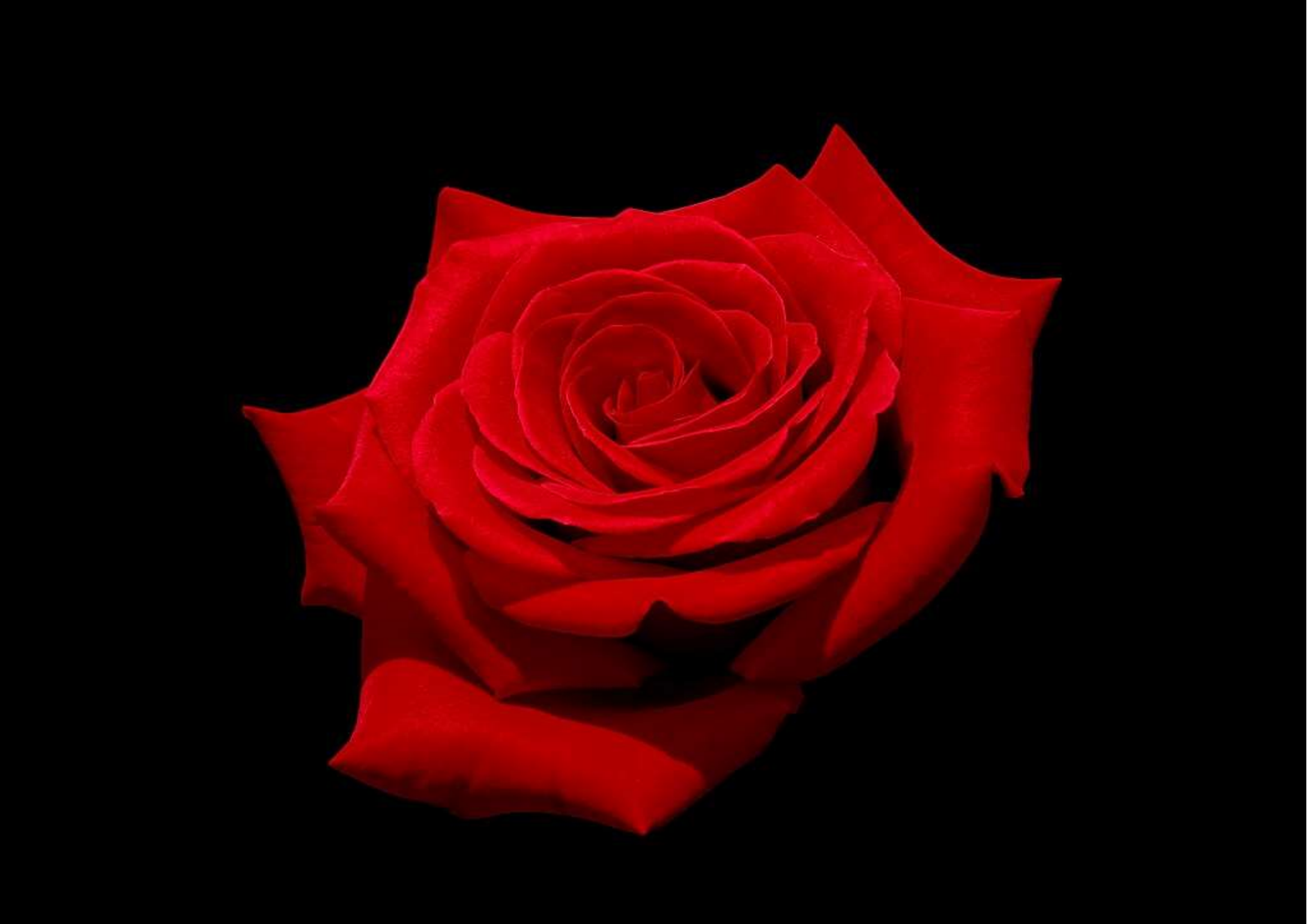}}
 \subfigure[]{\includegraphics[width=0.5\linewidth]{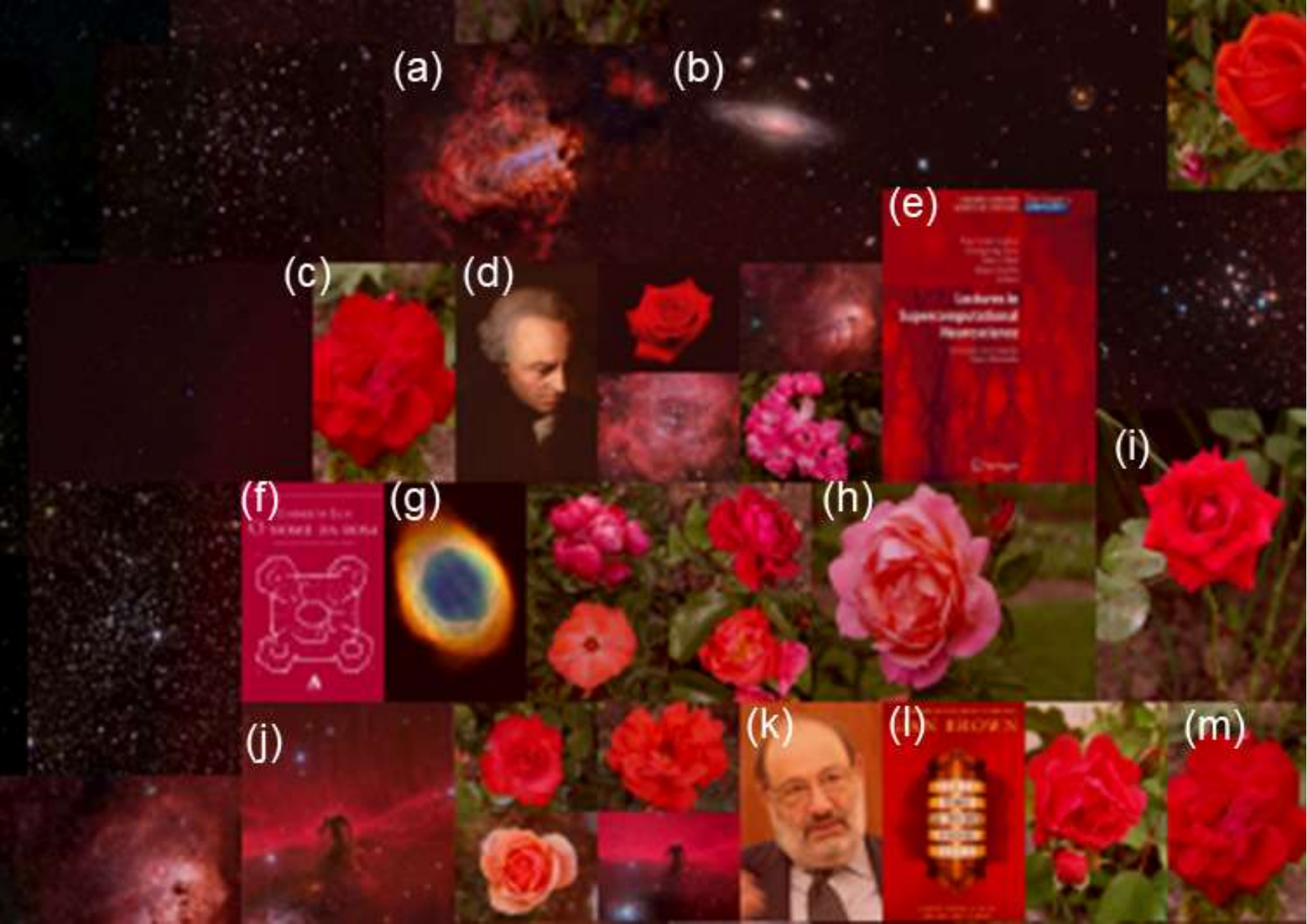}}
 \subfigure[]{\includegraphics[width=0.5\linewidth]{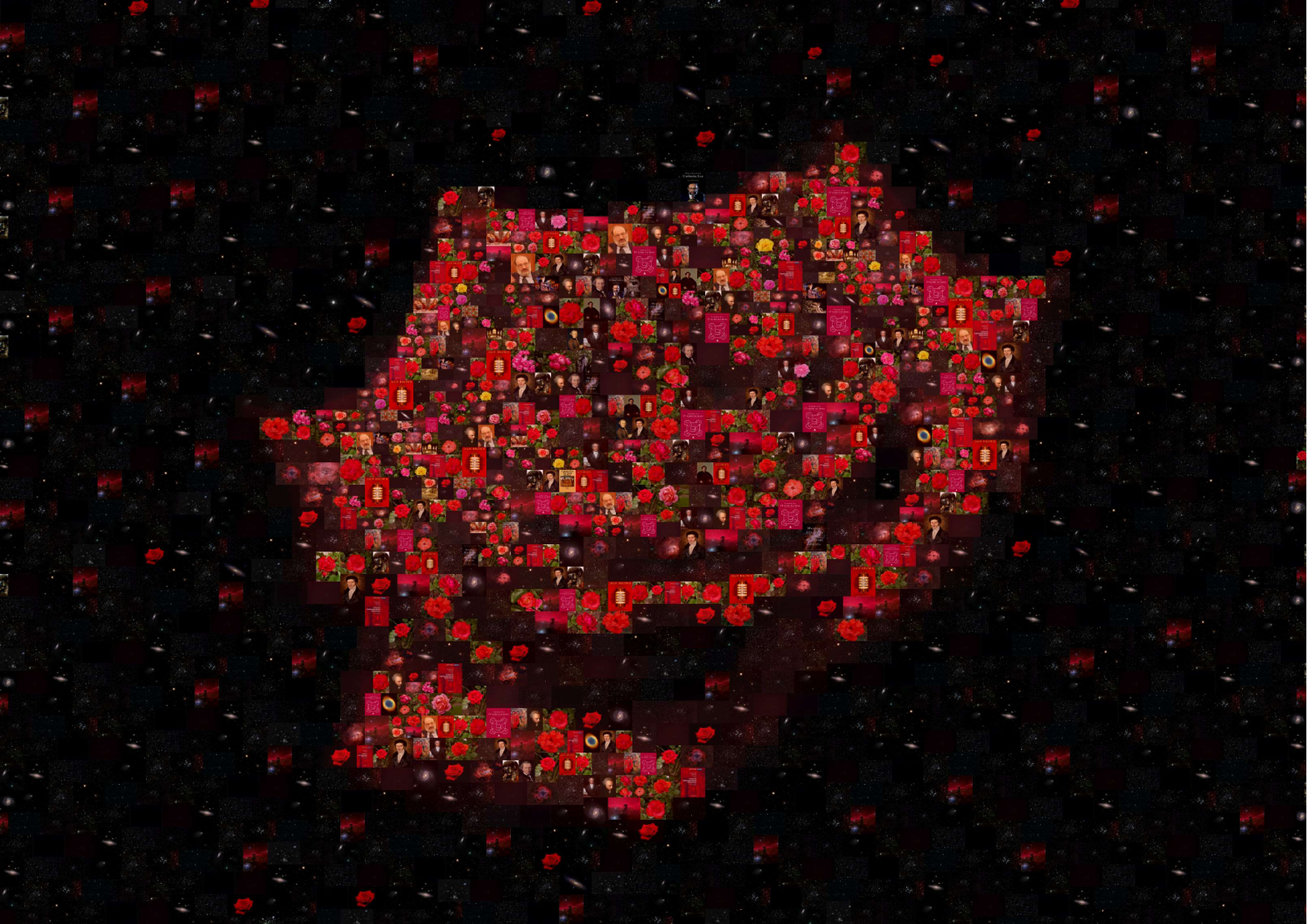}}
\caption{Rose associations. (a) Red rose with black background. (b) Image tiling. (c) Photomosaic of (a) with details shown in (b). (Color online)}\label{fig:roses}
\end{figure}

In order to present an intuitive illustration, I consider optic perception in \Fig{fig:roses}(a), instead of music, which is hardly to visualize. Figure \ref{fig:roses}(a) depicts a red rose with black background, as ``flowers are free natural beauties.'' \citep[§16, p.~81]{Kant14}.\footnote{
    Source: \url{https://commons.wikimedia.org/wiki/File:Red_rose_with_black_background.jpg}
}
Let $r$ be constant and $R$ be a predicate of first order logic such that $\mng{r} \in M$ (the meaning of $r$ in the domain $M$) refers to the object shown in \Fig{fig:roses}(a) and $\mng{R} \subseteq M$ (the meaning of $R$ in the domain $M$) is the subset of all roses contained in $M$.

Now, the effect of the \emph{determinant} power of judgement \citep[§IV, p.~17]{Kant14} can be easily identified with \emph{predication}. Subsuming an empirical object under a given concept, such as `the object in \Fig{fig:roses}(a) is a rose' expresses the model-theoretic relation
\begin{equation}\label{eq:predication}
  \mng{R(r)} = (\mng{r} \in \mng{R}) \:.
\end{equation}

In order to illustrate the \emph{reflective} power of judgement \citep[§IV, p.~17]{Kant14} in terms of model theory, consider \Fig{fig:roses}(b). This panel depicts a number of image tiles, some of them (e.g. tiles (c), (h), (i), (m) and many others) showing roses, while others presenting persons, astronomic objects or book covers, etc. (for image sources see Appendix).

One important function of the reflective power of judgement is the opposite of subsumption, namely unification:

\begin{quote}
The principle of reflection on given objects of nature is that for all things in nature empirically determinate \emph{concepts} can be found. \citet[p.~211]{Kant00}
\end{quote}
Where the ``reflective Judgement [\dots] is obliged to ascend from the particular in nature to the universal'' \citep[§IV, p.~18]{Kant14}. Now, given some rose-objects in \Fig{fig:roses}(b), the reflective power of judgement unifies these together under the concept label `roses', or, formally,
\begin{equation}\label{eq:predication}
  \mng{R} = \bigcup_{r: R(r)} \mng{r} \:.
\end{equation}

Yet, another eminent function of the reflective power of judgement besides unification is \emph{representation} (also called ``presentation'' or ``exhibition''):\footnote{
    Also cf.  \citet[p.~220]{Kant00})
}

\begin{quote}
    If the concept of an object is given, the business of the Judgement in the use of the concept for cognition consists in presentation (\emph{exhibitio}) i.e. in setting a corresponding intuition beside the concept. \citep[§VIII, p.~35]{Kant14}.
\end{quote}

For the aim of representation, the cognitive faculties, reflective power of judgement, understanding, and imagination have to play tightly together. In model-theoretic semantics, this interaction could be straightforwardly described by means of Zermelo's \emph{axiom of choice}, stating that for a non-empty family $\mathcal{P}$ of subsets $P \in \mathcal{P}, P \subseteq M$, there is a selection function $\sigma: \mathcal{P} \to M$, such that $\sigma(P) = m \in M$ \citep{Moore78}. Therefore, Kant's idea of (re-)presentation can be captured in the following way: Given a concept, i.e. a predicate, $R$ (e.g. `rose'), apply the selection function $\sigma$ to its extension,
\begin{equation}\label{eq:choice}
  \sigma(\mng{R}) = \mng{r} \in \mng{R}
\end{equation}
to present a characteristically representing image $\mng{r}$ of a rose in intuition.

Finally, these model-theoretic reconstructions can be brought together for illustrating Kant's key concept of the ``free play of the cognitive faculties'' \citep[§9, p.~64]{Kant14}. To this end, \emph{photomosaics} present a suitable mean of illustration \citep{Silvers96, Mizraji23}. A photomosaic is a computer-generated tiling of a given image such that the tiles are appropriately adjusted through their mean optical properties, such as color and brightness. Photomosaics share also some interesting properties with fractals such as the Mandelbrot set which exhibit their own aesthetic appeal \citep{PeitgenRichter86}. Clearly, by iteratively generating a photomosaic through recursive photomosaic tiles, would produce a fractal-like structure.

Consider \Fig{fig:roses}(c), depicting a photomosaic of the rose shown in \Fig{fig:roses}(a), that I have created with \emph{AndreaMosaic}\footnote{
    \url{http://www.andreaplanet.com/andreamosaic/}
}
with \Fig{fig:roses}(b) as a zoom view into its details. Most of the tiles within the rose domain present different kinds of roses, altogether constituting the extension of the empirical concept `rose'. By contrast, the black background is tiled with dark images of some astronomical objects.

Looking at the details in \Fig{fig:roses}(b), reveals a clipping of my personal semantic web: I am writing an article about Kant's philosophical aesthetics, therefore tile (d) shows a picture of Kant in the very same year when he published the \emph{Critique of the Power of Judgement}. There, Kant discusses flowers as examples of ``free natural beauties'' \citep[§16, p.~81]{Kant14}. The concept of `flower' is a unification of the concept of `rose', hence, tiles (c), (i), (h), (m) present particular roses in my imagination. In the \emph{Critique of Practical Reason}, \citet[p.~162]{Kant15} talked about ``the starry heavens above me'' as one object of ``increasing admiration and reverence''. Thus, I have selected images (a), (b),  (j), and (g) as representatives for heavenly objects. Thinking about roses in literary art, I am further associating \emph{The Name of the Rose} (tile f) of Italian semiotician Umberto Eco (tile k), but also Dan Brown because of his ``sub rosa'' leitmotif in the \emph{Da Vinci Code} (tile l). Finally, my mind is wandering back to my own scientific work, as indicated by tile (e). In different regions of the photomosaic \Fig{fig:roses}(c), other related associations have taken place: e.g. there is one tile showing Gertrude Stein: ``a rose is a rose [\dots]'' and another one with the book cover of music psychologist Diana Deutsch's \emph{Musical Illusions and Phantom Words}, etc. (not shown).

Thus, during my reception of the photomosaic \Fig{fig:roses}(c), I am experiencing a kind of ``free play'' \citep[§9, p.~64]{Kant14} of my own intuitive imagination, though under the ``laws of association'' \citep[§29, p.~136]{Kant14} imposed by my understanding of concepts. This lawfulness play without a law somehow appears ``harmonious'' \citep[§39, p.~168]{Kant14}, thereby conveying a feeling of pleasure to my soul \citep[§15, p.~80]{Kant14}. Finally, I am judging the mosaic \Fig{fig:roses}(c) (but also the rose \Fig{fig:roses}(a)) `beautiful', not, because `beauty' can directly be attributed to the rose, but because it is the reason for my pleasure in the harmonious interaction between imagination and understanding \citep[§12, p.~71]{Kant14}. Now, employing an idea of \citet[p.~138]{Huron06}, the aesthetic judgement comprises a \emph{missattribution} from the source of the feeling of pleasure, the harmony of the faculties, to its stimulus, the given `beautiful' object.\footnote{
Compare:
\begin{quote}
    This Deduction is thus easy, because it has no need to justify the objective reality of any concept, for Beauty is not a concept of the Object and the judgement of taste is not cognitive. \citep[§38, p.~166]{Kant14}.
\end{quote}
See also the thorough discussion by \citet{Blutner24b}.
}


\subsection{Generative Adversarial Network Theory on Kant}
\label{sec:gantkant}

Another crucial aspect of the reflective power of judgement refers to the situation ``if only the particular be given for which the universal has to be found'' \citep[§IV, p.~ 17]{Kant14}. This is generally the case for \emph{concept formation}, \emph{categorization}, and \emph{learning} \citep{RussellNorvig10}. By identifying concepts with the extensions of predicates, every predicate $P$ together with its logical negation $\neg P$, defines a binary partition $\mathcal{P} = \{A_1, A_2 \}$ of the model-theoretic discourse domain $M = A_1 \cup A_2$, such that $A_1 = \mng{P}$ and $A_2 = \mng{\neg P}$.  (Likewise, a family of disjoint concepts, e.g. color terms, provides a non-binary partition of $M$). Thus, concept formation turns out as clustering through the acquisition of partition boundaries \citep{RussellNorvig10}.

In machine learning approaches, discourse domains are usually codified as high-dimensional feature spaces $X \subseteq \mathbb{R}^n$, with dimensionality $n \in \mathbb{N}$ \citep{RussellNorvig10}. In vision science, e.g., a grey scale image such as the rose in the hard copy version of \Fig{fig:roses}(a), that was discretized into $n = 1024 \times 787 = 805,888$ pixels, each one contributing a real value ranging from black (=0) to white (=1), becomes represented by a vector $\vec{x} \in X = [0, 1]^n \subseteq \mathbb{R}^n$. Correspondingly, an acoustic spectrogram can be regarded as a real-valued image in the dimensions of time $\times$ frequency, leading to a similar vectorial sampling. Such high-dimensional images have to be compressed by virtue of advanced data analysis techniques (such as principal component analysis, hidden Markov models, etc) in order to obtain a suitable feature space for subsequently employing machine learning techniques \citep{RussellNorvig10}.

An important class of classifiers are \emph{perceptrons}, which are one- or multi-layered feed-forward neural networks with nonlinear activation functions \citep{HertzKroghPalmer91, RussellNorvig10}. Let $\vec{x} \in X$ be an $n$-dimensional input vector and $\vec{y} \in Y \subseteq \mathbb{R}^m$ be an $m$-dimensional output vector, such that admissible outputs are restricted to $y_i = 1$ for one output unit $i$ and $y_k = 0$ for all other output units $k \ne i$. Then, the neural network equation
\begin{equation}\label{eq:percept}
  \vec{y} = \Theta( \vec{W} \cdot \vec{x})
\end{equation}
with Heaviside step activation function $\Theta(x) = 1(0)$ if $x \ge 0 (<0)$, and synaptic weight matrix $\vec{W}$ could describe a one-layered perceptron that works as a linear classifier separating several data clusters in input space though linear partition boundaries where the activation function acts as a decision function over some decision thresholds that are encoded as biases in the weight matrix \citep{HertzKroghPalmer91, RussellNorvig10}. Training such a network by means of the perceptron learning rule (a simplification of the multi-layered backpropagation rule) leads to the emergence of average prototype representations in input space,\footnote{
    Interestingly, Kant already speculated about an ``archetype of beauty'' \citep[§17, p.~88]{Kant14} produced by a kind of averaging algorithm:

    \begin{quote}
    Further, if the mind is concerned with comparisons, the Imagination can, in all probability, actually though unconsciously let one image glide into another, and thus by the concurrence of several of the same kind come by an average, which serves as the common measure of all.  \citep[§17, p.~87]{Kant14}
    \end{quote}

    This idea has been empirically validated for a first time by \citet{Galton1878}; cf. the more recent review by \citet{Collins12}. Also note the close connection to matrix memory neural networks \citep{Mizraji10}.
}
from which classification results are obtained through minimizing the distances of an actual input vector to the respective class prototypes. This method underlies, e.g., the key-finding algorithm of \citet{Krumhansl90} in music retrieval.

Replacing the Heaviside function in \eqref{eq:percept} by the logistic activation function
\begin{equation}\label{eq:logact}
    f(x) = \frac{1}{1 + \E^{-x}} \:,
\end{equation}
converts the hard decision problem into a soft, probabilistic, one such that $f(x) = \Pr(x \ge 0)$ becomes the probability that input $x$ belongs to a given class \citep{HertzKroghPalmer91, RussellNorvig10}.

A binary classifier with a single output unit $y$ is called \emph{discriminator}. For a probabilistic discriminator, the input space $X$ is partitioned into two categories $X = A_1 \cup A_2$, corresponding to the decisions that $\vec{x} \in A_1$ or $\vec{x} \not\in A_1$ when $A_2 = X \setminus A_1$ is the complement of class $A_1$. Then,
\begin{equation}\label{eq:discrim}
  y = D(\vec{x}) = \Pr(\vec{x} \in A_1) = \Pr(A_1 | \vec{x})
\end{equation}
is the (conditional) probability that the input $\vec{x}$ is correctly classified as a member of $A_1$. Correspondingly, $1 - D(\vec{x}) = \Pr(\vec{x} \not\in A_1)$ becomes the probability of the converse classification problem that $\vec{x}$ belongs to $A_2$, instead.

\begin{figure}[H]
\centering
\includegraphics[width=0.9\linewidth]{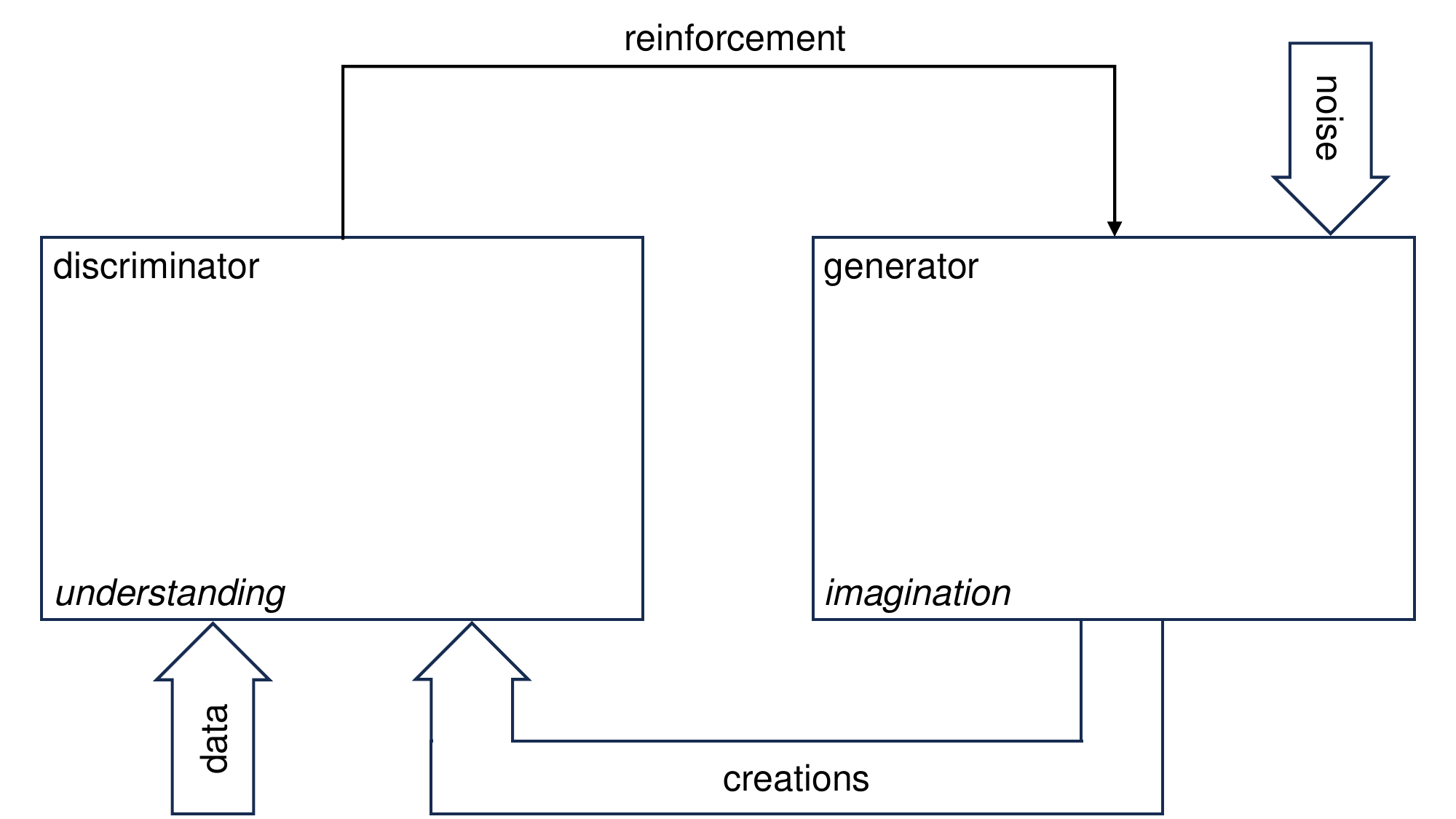}
\caption{Architecture of a generative adversarial network (GAN).}\label{fig:gan}
\end{figure}

Now, a generative adversarial network (GAN) is a (deep) neural network, where two modules \citep{CarmantiniEA17}, a discriminator and another one, called \emph{generator}, are recurrently coupled together in a \emph{reinforcement} loop \citep{GoodfellowEA14}. Figure \ref{fig:gan} presents its architecture schematically, where both modules are indicated as `black boxes' comprising quite different neural topologies, ranging from multi-layered perceptrons \citep{GoodfellowEA14}, to continuous recurrent neural networks (C-RNN) \citep{Mogren16} or convolutional neural networks (CNN) \citep{YangChouYang17}, as used for AI music composition.

The discriminator in \Fig{fig:gan} is trained on two different data sets, natural examples $\vec{x}$, drawn from a big data pool (\emph{data}) and \emph{creations} $\vec{y}$ delivered by the generator module. Both inputs must have the same dimensionality, $\vec{x},  \vec{y} \in X \subseteq \mathbb{R}^n$, but are initially governed by two different probability distribution density functions: $\vec{x} \sim \rho$ and $\vec{y} \sim \gamma$, with $\rho, \gamma : X \to \mathbb{R}$. The discriminator  returns a single output $D(\vec{x}) \in \mathbb{R}$ which is the probability that the input vector $\vec{x}$ belongs to the class of natural examples $A_1$, according to \eqref{eq:discrim}.

The output of the discriminator module is fed into the generator module as a reinforcement signal from which the generator can determine its actual reward or penalty. The purpose of the generator is to hocus the discriminator by creating outputs that cannot be distinguished from the natural example data. Thus, the generator may apply a simple thresholding algorithm: if $D(\vec{y}) > 0.5$ for its own output $\vec{y}$ the discriminator has successfully been fooled because the generated datum $\vec{y}$ had mistakenly been classified as natural $\vec{y} \in A_1$ by the discriminator, thereby leading to the generator's reward. If, on the other hand,  $D(\vec{y}) < 0.5$, the discriminator was successful in telling the difference between the generated datum $\vec{y}$ and a natural representative, which leads to the generator's punishment.

More specifically, the generator is able to freely create its output by filtering some random noise signal $\vec{z} \in X$ from its input. Under the assumption that the noise has another probability distribution density function $\nu : X \to \mathbb{R}$, the action of the generator can mathematically be described by a transfer function $G : X \to X$, such that $\vec{y} = G(\vec{z})$, implementing a Frobenius-Perron operator \citep{Ott93, TrollGraben98}
\begin{equation}\label{eq:fpo}
  \gamma(\vec{x}) = \EV_{\vec{z} \sim \nu} [\delta(\vec{x} - G(\vec{z}))]
\end{equation}
with Dirac's delta distribution as integration kernel in the expectation value functional $\EV$ for the random variable $\vec{z}$ drawn from the noise distribution $\nu$.

The training objective of the GAN is then formalized by the temporal limit
\begin{equation}\label{eq:ganaim}
  \gamma \stackrel{T \to \infty}{\longrightarrow} \rho \:,
\end{equation}
i.e., the output distribution of the generator approaches the distribution of the natural training data for increasing training time $T$. This is achieved by reinforcement learning \citep{Williams88, HertzKroghPalmer91, Schmidhuber10a, Schmidhuber12} through an \emph{adversarial  minimax game}, played by the interacting modules \citep{RussellNorvig10}.

To this end, one defines a \emph{utility function} of the discriminator as
\begin{equation}\label{eq:disutil}
  U(D, G) = \EV_{\vec{x} \sim \rho} [\log(D(\vec{x}))] + \EV_{\vec{z} \sim \nu} [\log(1 - D(G(\vec{z})))]
\end{equation}
where is first term denotes the discriminator's reward from correctly classifying a natural example $\vec{x} \in X$, while the second term describes the rewarding contribution from correctly rejecting a creation $\vec{y} = G(\vec{z})$ of the generator. Correspondingly, a \emph{cost function} of the generator
\begin{equation}\label{eq:gencost}
  C(D, G) = \EV_{\vec{z} \sim \nu} [\log(1 - D(G(\vec{z})))]
\end{equation}
is given by the second term of the discriminator's utility function, since a correct rejection of a generator's creation $\vec{y} = G(\vec{z})$ by the discriminator is penalized as a failed attempt to fool the latter module. Hence, the GAN training objective is expressed as
\begin{equation}\label{eq:ganopt2}
  \min_G \max_D \EV_{\vec{x} \sim \rho} [\log D(x)] + \EV_{\vec{z} \sim \nu} [\log (1 - D(G(z)))]
\end{equation}
which is essentially GAN's signature \Eq{eq:ganopt} from the auction at \citet{Christies18}.

Provided that the GAN architecture \Fig{fig:gan} has unlimited resources, \citet{GoodfellowEA14} were able to prove the convergence \eqref{eq:ganaim} of the training algorithm. One important step in their proof comprises the statement that for a given generator transfer function $G$ and its corresponding probability density function $\gamma$, there is one optimal discriminator function $D^*$, that maximizes the utility \eqref{eq:disutil}, entailing
\begin{equation}\label{eq:optutil}
  D^*(\vec{x}) = \frac{\rho(\vec{x})}{\rho(\vec{x}) + \gamma(\vec{x})} \:.
\end{equation}
Then, the discriminator's utility function can be expressed either by Kullback-Leibler divergences, or, likewise, by its symmetrical counterpart, called Jensen-Shannon divergence \citep{GoodfellowEA14}, which relates its optimization to Bayesian exploration \citep{IttiBaldi05, SunGomezSchmidhuber11} and pragmatic information theory in dynamic semantics \citep{Graben06}. In the limit \eqref{eq:ganaim}, when the generator's distribution reproduces the natural data distribution, $\gamma = \rho$, \Eq{eq:optutil} yields $D^*(\vec{x}) = 0.5$, i.e. the generator outperforms the discriminator, that is thereby acting at chance level.

Next, I supply the isomorphism between Kant's analytic of the beautiful, outlined in the \emph{Critique of the Power of Judgement} and generative adversarial network theory (GANT) \citep{GoodfellowEA14, Schmidhuber10a, Schmidhuber12}. To this end, I subsequently discuss a few iterations of the learning algorithm.

At initialization, it could be assumed that the untrained generator acts as the identity operator $G(\vec{z}) = \vec{z}$ upon the random noise input $\vec{z} \in X$. Hence, the generator's output distribution simply reproduces the noise distribution $\gamma(\vec{z}) = \nu(\vec{z})$. Since the discriminator is also untrained at initialization, it merely classifies at chance level, $D(\vec{x}) = 0.5$. After some iterations of batch learning, the discriminator approaches its relative optimum \Eq{eq:optutil}, $D^*(\vec{x}) = \rho(\vec{x}) / (\rho(\vec{x}) + \nu(\vec{x}))$, therefore becoming able to correctly predict and classify the given natural training vectors.\footnote{
    Note that \citet[p.~233]{Schmidhuber10a} emphasized that ``there is a deep connection between optimal prediction and optimal compression'', where compression is clearly related to classification.
}
After another round of batch iterations, also the generator has improved its performance hocussing the discriminator from trial to trial. Both modules are engaged in their adversarial game until convergence, when the discriminator is degraded to change level again.

My reconstruction of Kant's theory of art reception in terms of GANT is then elucidated in \Tab{tab:kantgan}.

\begin{table}[H]
  \centering
  \small
  \begin{tabular}{|*{3}{l|}}
    \hline
    Kant & GANT & \citet{Kant14} \\
    \hline
    ``feeling of pleasure and pain'' & reward and punishment & §III, p.~15 \\
    ``determinant'' Judgement  & predication, discrimination & §IV, p.~ 17 \\
    ``reflective'' Judgement & prediction, learning & §IV, p.~ 17 \\
    ``Understanding, as the faculty of concepts'' & discriminator & §VII, p.~31 \\
    ``Imagination (as a productive faculty of cognition)'' &  generator & §49, p.~197 \\
    ``free play'' & adversarial game & §9, p.~64 \\
    ``harmony of the cognitive faculties'' & algorithm's convergence & §9, p.~65 \\
    ``spontaneity'' &  random forcing & §IX, p.~41 \\
    ``purposiveness without purpose'' & generator terminal & §22, p.~96 \\
    ``law[fulness] without a law'' & discriminator terminal & §22, p.~96 \\
    ``pretty[iness]'' & prototypicality & §49, p.~197 \\
    ``beauty'' & interestingness & §39, p.~168 \\
    ``furtherance of the whole life'' &  training objective & §54, p.~221 \\
    \hline
  \end{tabular}
\caption{Isomorphism between the aesthetic model of Kant's \emph{Critique of the Power of Judgement} and generative adversarial network theory (GANT).}\label{tab:kantgan}
\end{table}

My basic idea for the reconstruction is that during art reception, the human mind acts in analogy to a GAN in its training phase while being stimulated by the many different episodes or views upon a given object. On the one hand, I identify understanding as the faculty of concepts with the GAN discriminator [\Fig{tab:kantgan}:~\emph{understanding}] which acts as a binary classifier. During GAN training the discriminator acquires law-like categories through the successful prediction and classification of natural input patterns. This operation mode is related to the reflective power of judgement. After training, the discriminator behaves in analogy to the determinant power of judgement by subsuming input patterns, either natural ones from the training pool or those ones created by the generator, under learned categories through predication.

On the other hand, I equate the productive faculty of imagination with the GAN generator [\Fig{tab:kantgan}:~\emph{imagination}]. Its capability to spontaneously create `art works' is described by the random forcing of the module in the GAN architecture. Both modules are engaged in an adversarial minimax game during the training phase, thus reflecting the free play of the cognitive faculties. As the generator aims at fooling the discriminator, which in turn, tries to make successful predictions about the source of a given input, the game develops towards an optimum, prescribed by the objective function \eqref{eq:disutil}, eventually accounting for the harmony of the algorithm \eqref{eq:ganopt2}.\footnote{
    Note that a particular approach in computational neuroscience is explicitly dubbed \emph{harmony theory} \citep{Smolensky86, SmolenskyLegendre06a, Smolensky06}.
}
Depending on the respective harmony pay-off, both modules are either rewarded or penalized within their reinforcement loop, which causes the feelings of pleasure and pain in a somewhat anthropocentric metaphor.

As soon as the convergence limit is approached, the discriminator becomes increasingly unable to distinguish the origin of its input patterns; thus it terminates in a state of lawfulness (after successfully acquiring many categories) without a law (operating at change level). By contrast, the generator has reached optimal purposiveness (deceiving the discriminator), but without any substantial purpose.

Moreover, my neural reconstruction allows to differentiate between prettiness and beauty according to Kant's formulations. For Kant, prettiness is subjectively appealing without the ``spirit'' of aesthetical ideas, which ``is the name given to the animating principle of the mind'' (\citet[§49, p.~197]{Kant14}; cf. also \citet[p.~66]{Kant11}) which has to be interpreted as the free, harmonious, and autonomous play of imagination and understanding. For a judgement of prettiness essentially lacks this autonomy of the cognitive faculties, I interpret prettiness as prototypicality in the sense of the ``archetype of beauty'' \citep[§17, p.~88]{Kant14}. Contrastingly, the judgement of the beautiful resides in mere ``reflection [of] the accordance of the representation with the harmonious (subjectively purposive) activity of both cognitive faculties in their freedom, i.e. to feel with pleasure the mental state produced by the representation.'' \citep[§39, p.~168]{Kant14}.\footnote{
    Note that my reconstruction crucially deviates in this point from the terminology of \citet{Schmidhuber10a}:

    \begin{quote}
      The subjective simplicity or compressibility or regularity or beauty [\dots] The observer-dependent and time-dependent subjective interestingness or surprise or aesthetic value is the first derivative of subjective simplicity [\dots] \citep[p.~234]{Schmidhuber10a}
    \end{quote}

    Therefore, Schmidhuber's notion of beauty corresponds to my (and Kant's) concept of prettiness, namely prototypicality as measured by compressibility; while his notion of subjective interestingness better corresponds with my (and Kant's) concept of beauty (cf. also the discussion in \citet{Collins12}).
}

Finally, my reconstruction is completed by the observation that aesthetic reception has an overall positive impact upon the life of the perceiver through the ``feeling of the furtherance of life'' \citep[§23, p.~102; §49, p.~201; §54, p.~221]{Kant14}. This aspect had also been emphasized by \citet{Schmidhuber10a, Schmidhuber12}:

\begin{quote}
  The current \emph{intrinsic} reward, \emph{creativity} reward, \emph{curiosity} reward, \emph{aesthetic reward}, or \emph{fun} $r_\mathrm{int}(t)$ of the action selector is the current \emph{surprise} or \emph{novelty} measured by the improvements of the world model $p$ at time $t$. \citep[p.~232]{Schmidhuber10a} (Italics in the original)
\end{quote}
Where the \emph{action selector} is another term for the generator, while the \emph{world model} denotes the GAN's discriminator, here.

Thus, a cognitive agent comprised by two GAN modules, discriminator/understanding and generator/imagination that is continuously engaged in curious interaction with the challenges of the world is permanently improving its prediction and classification capability and seldom becomes bored \citep{KenettHumphriesChatterjee23}. This is, of course, also of particular significance for the scientific endeavor when the \emph{eureka effect} emotionally rewards the scientist for the rational unification of diverse empirical laws \citep{Stangneth19, Mizraji23}:

\begin{quote}
  Hence, as if it were a lucky chance favouring our design, we are rejoiced (properly speaking, relieved of a want), if we meet with such systematic unity under merely empirical laws; although we must necessarily assume that there is such a unity without our comprehending it or being able to prove it. \citep[§V, p.~24]{Kant14}
\end{quote}

The latter quotation provides the connection of Kant's aesthetics with his theory of teleology that is required as a regulative principle for the unification of science \citep{Kant14, Primas90a}.


\section{Discussion}
\label{sec:disc}

In this study, I have presented a reconstruction of Kant's theory of art reception in terms of artificial intelligence (AI) \citep{GoodfellowEA14, Schmidhuber10a, Schmidhuber12}. More specifically, I have used the theory of generative adversarial networks (GANT) for demonstrating an isomorphism between both frameworks in case of art reception. The central pillar of my approach is the analogy between the discriminator module of a GAN, capturing the lawfulness of its perceptions with Kant's understanding as the faculty of rules \citep[B171]{Kant99} on one side, and the GAN generator module with Kant's productive imagination (\citet[B181]{Kant99}, \citet[§49, p.~198]{Kant14}) as the faculty of intuition \citep[§39, p.~168]{Kant14}. During network training, both neural modules are engaged in a free adversarial game, with the generator attempting to fool the discriminator, which, in turn, is constantly improving its discrimination performance. The interaction of both modules is harmonious approaching convergence, which is felt as subjective pleasure in Kant's account \citep[§12, p.~71]{Kant14}. Referring to an idea of \citet[p.~138]{Huron06}, this pleasure becomes missattributed from its source to its stimulus, i.e. the beautiful object in a judgement of taste \citep[§38, p.~166]{Kant14}.

In order to probe the suggested reconstruction, I discuss a recent experiment of \citet{CheungHarrisonEA19} on empirical music aesthetics. They reported high ratings of aesthetic appreciation for both, either predictable closures in rather uncertain contexts, and for surprising closures in relatively certain contexts. These findings are essentially consistent with my GANT approach for the following reasons. First, consider the case of an uncertain context. In this setting, the discriminator has not sufficiently been trained to capture all the regularities of the natural data pool. However, if the discriminator is yet able to successfully classify an unexpected creation produced by the generator, the discriminator wins its match against the generator, thus being rewarded by the second term of the utility function \Eq{eq:disutil}. Secondly, in the other case of a relatively certain context, the discriminator is close to its optimum \Eq{eq:optutil} for a given generator distribution. If now the generator is able to successfully hocus the discriminator, the latter becomes surprised by a novel creation of the former, that consequently wins this match against the discriminator. Then, this outcome is rewarding for the generator. Hence, in both cases, the GAN modules are in an harmonious interplay, as described by the minimax algorithm \Eq{eq:ganopt2}, resulting in high aesthetic appreciation. To summarize this argument, a listener who is familiar with a particular musical style could easily predict the temporal patterns of an unknown piece conforming to that given style. In such a situation of certainty, an unpredictable musical event is surprising and signals that the discriminator has not yet approached its optimum. Thus, there is a need for cognitive improvement. In the contrasting uncertain situation, the listener is not sufficiently experienced to make appropriate predictions. Though, when she is yet able to systematically predict musical events beyond change level, her generator has also not reached its optimum for it could be further improved to fool the discriminator. In both cases, the demand to further improve the mental faculties for aesthetic reception is felt as pleasure.

Finally, I address the question whether and to which extent generations of AI could count as \emph{real art}. In an interview with Christie's auction house, the director of the Art and Artificial Intelligence Lab at Rutgers University, Ahmed Elgammal, said:

\begin{quote}
  Yes, if you look just at the form, and ignore the things that art is about, then the algorithm is just generating visual forms and following aesthetic principles extracted from existing art. But if you consider the whole process, then what you have is something more like conceptual art than traditional painting. There is a human in the loop, asking questions, and the machine is giving answers. That whole thing is the art, not just the picture that comes out at the end. You could say that at this point it is a collaboration between two artists --- one human, one a machine. And that leads me to think about the future in which AI will become a new medium for art. \citep{Christies18}
\end{quote}

Neglecting some ethical issues about origin, bias, and diversity of AI training data \citep{BenderGebruEA21},\footnote{
    Letter signed by more than 200 artists makes broad ask that tech firms pledge to not develop AI tools to replace human creatives: \url{https://www.theguardian.com/technology/2024/apr/02/musicians-demand-protection-against-ai}
}
AI applications could lead, according to Elgammal, to new forms of artistic activity that might be called \emph{AI assisted arts (AIAA)}. What does this imply for my suggested GANT reconstruction? First of all, all mathematical proofs on the convergence of the GAN algorithm require infinite resources such as memory capacity and cardinality of the natural training data \citep{GoodfellowEA14, Schmidhuber10a, Schmidhuber12}. This is clearly not ensured for contemporary AI systems. Yet this fact has immediate consequences for the understanding of artificial artists. Following Kant:

\begin{quote}
  The mental powers, therefore, whose union (in a certain relation) constitutes genius are Imagination and Understanding [\dots] \emph{Thus genius properly consists in the happy relation [between these faculties], which no science can teach and no industry can learn}, by which Ideas are found for a given concept [\dots] The latter talent is properly speaking what is called spirit [\dots] (\emph{which is even on that account original and discloses a new rule that could not have been inferred from any preceding principles or examples}), that can be communicated without any constraint [of rules]. \citep[§49, p.~201]{Kant14} (My italics)
\end{quote}

Here, the crucial idea is that becoming a genius cannot be learned by obeying prescribed artistic rules. By contrast, genius is able to mold new rules. Moreover:

\begin{quote}
  In accordance with these suppositions \emph{genius is the exemplary originality of the natural gifts of a subject in the free employment of his cognitive faculties. In this way the product of a genius (as regards what is to be ascribed to genius and not to possible learning or schooling) is an example, not to be imitated (for then that which in it is genius and constitutes the spirit of the work would be lost), but to be followed, by another genius; whom it awakens to a feeling of his own originality and whom it stirs so to exercise his art in freedom from the constraint of rules, that thereby a new rule is gained for art, and thus his talent shows itself to be exemplary. But because a genius is a favourite of nature and must be regarded by us as a rare phenomenon, his example produces for other good heads a school, i.e. a methodical system of teaching according to rules, so far as these can be derived from the peculiarities of the products of his spirit. For such persons beautiful art is so far imitation, to which nature through the medium of a genius supplied the rule.} \citep[§49, p.~203f]{Kant14} (My italics)
\end{quote}

Therefore, the power of judgement can only be trained by means of examples; and a genius is able to create examples of art, from which other geniuses may form their artistic productivity. Finally:

\begin{quote}
    Abundance and originality of Ideas are less necessary to beauty than the accordance of the Imagination in its freedom with the conformity to law of the Understanding. \emph{For all the abundance of the former produces in lawless freedom nothing but nonsense}; [\dots] \citep[§50, p.~205]{Kant14} (My italics)
\end{quote}

Hence, according to Kant's theory of art generation, contemporary AI artists may not be called \emph{real geniuses} for they are trained by means of finite examples to either reproduce particular styles, or to produce some kind of novelty. In the first case, AI art may be called \emph{manneristic}, in the second merely \emph{random} \citep{Zoeller24}.


\section*{Acknowledgements}

I gratefully acknowledge financial support by the German Federal Agency of Labor. I also thank Eduardo Mizraji, Matthias, Zoeller, Serafim Rodrigues, and Reinhard Blutner for inspiring insights.

\vspace{3cm}\hspace*{\fill}\begin{minipage}[b]{7cm}
\footnotesize
Wie fang ich nach der Regel an? --- \\
Ihr stellt sie selbst und folgt ihr dann. \\
Gedenkt des sch\"onen Traums am Morgen; \\
f\"urs andre la\ss{}t Hans Sachs nur sorgen!
Richard Wagner, \emph{Die Meistersinger von N\"urnberg} (1868),
3rd Act, 2nd Scene
\end{minipage}


\bibliographystyle{apalike}

\nocite{Kukla06}


\begin{thebibliography}{}

\bibitem[Achourioti and van Lambalgen, 2011]{AchouriotiLambalgen11}
Achourioti, T. and van Lambalgen, M. (2011).
\newblock {A formalization of Kant’s transcendental logic}.
\newblock {\em Review of Symbolic Logic}, 4(2):254 -- 289.

\bibitem[beim Graben, 2006]{Graben06}
beim Graben, P. (2006).
\newblock Pragmatic information in dynamic semantics.
\newblock {\em Mind and Matter}, 4(2):169 -- 193.

\bibitem[beim Graben, 2014]{Graben14a}
beim Graben, P. (2014).
\newblock Order effects in dynamic semantics.
\newblock {\em Topics in Cognitive Science}, 6(1):67 -- 73.

\bibitem[Bender et~al., 2021]{BenderGebruEA21}
Bender, E.~M., Gebru, T., McMillan-Major, A., and Shmitchell, S. (2021).
\newblock On the dangers of stochastic parrots: Can language models be too big?
\newblock In {\em Proceedings of the ACM Conference on Fairness,
  Accountability, and Transparency}, pages 610 -- 623, New York (NY). ACM.

\bibitem[Blutner, 2024]{Blutner24b}
Blutner, R. (2024).
\newblock The emotional meaning of pure music.
\newblock {\em Mind and Matter}, this issue.

\bibitem[Briot et~al., 2020]{BriotHadjeresPachet20}
Briot, J.-P., Hadjeres, G., and Pachet, F.-D. (2020).
\newblock {\em Deep Learning Techniques for Music Generation}.
\newblock Computational Synthesis and Creative Systems. Springer, Cham.

\bibitem[Carmantini et~al., 2017]{CarmantiniEA17}
Carmantini, G.~S., beim Graben, P., Desroches, M., and Rodrigues, S. (2017).
\newblock A modular architecture for transparent computation in recurrent
  neural networks.
\newblock {\em Neural Networks}, 85:85 -- 105.

\bibitem[Cheung et~al., 2019]{CheungHarrisonEA19}
Cheung, V. K.~M., Harrison, P. M.~C., Meyer, L., Pearce, M.~T., Haynes, J.-D.,
  and Koelsch, S. (2019).
\newblock Uncertainty and surprise jointly predict musical pleasure and
  amygdala, hippocampus, and auditory cortex activity.
\newblock {\em Current Biology}, 29(23):4084 -- 4092.e4.

\bibitem[Christie's, 2018]{Christies18}
Christie's (2018).
\newblock Is artificial intelligence set to become art’s next medium?
\newblock Online:
  \url{https://www.christies.com/en/stories/a-collaboration-between-two-artists-one-human-one-a-machine-0cd01f4e232f4279a525a446d60d4cd1}.
\newblock Retrieved at Saturday, April, 6th, 2024.

\bibitem[Collins, 2012]{Collins12}
Collins, M. (2012).
\newblock The attractiveness of the average face.
\newblock {\em Seminars in Orthodontics}, 18(3):217 -- 228.

\bibitem[Galton, 1878]{Galton1878}
Galton, F. (1878).
\newblock Composite portraits.
\newblock {\em Nature}, 18(447):97 -- 100.

\bibitem[G\"ardenfors, 1988]{Gardenfors88}
G\"ardenfors, P. (1988).
\newblock {\em Knowledge in Flux. Modeling the Dynamics of Epistemic States}.
\newblock MIT Press, Cambridge (MA).

\bibitem[Gatys et~al., 2016]{GatysEckerBethge16}
Gatys, L.~A., Ecker, A.~S., and Bethge, M. (2016).
\newblock Image style transfer using convolutional neural networks.
\newblock In {\em Proceedings of the IEEE Conference on Computer Vision and
  Pattern Recognition (CVPR)}.

\bibitem[Gibbs, 2022]{Gibbs22}
Gibbs, A.~C. (2022).
\newblock Aesthetics of music, on psychedelics.
\newblock {\em Mind and Matter}, 20(2):177 -- 194.

\bibitem[Ginsborg, 1997]{Ginsborg97}
Ginsborg, H. (1997).
\newblock {Lawfulness without a law: Kant on the free play of imagination and
  understanding}.
\newblock {\em Philosophical Topics}, 25(1):37 -- 81.

\bibitem[Ginsborg, 2003]{Ginsborg03}
Ginsborg, H. (2003).
\newblock Aesthetic judging and the intentionality of pleasure.
\newblock {\em Inquiry}, 46(2):164 -- 181.

\bibitem[Goodfellow et~al., 2014]{GoodfellowEA14}
Goodfellow, I., Pouget-Abadie, J., Mirza, M., Xu, B., Warde-Farley, D., Ozair,
  S., Courville, A., and Bengio, Y. (2014).
\newblock Generative adversarial nets.
\newblock In Ghahramani, Z., Welling, M., Cortes, C., Lawrence, N., and
  Weinberger, K.~Q., editors, {\em Advances in Neural Information Processing
  Systems (NIPS)}, volume~27. Curran Associates, Inc.

\bibitem[Guyer, 2006]{Guyer06b}
Guyer, P. (2006).
\newblock The harmony of the faculties revisited.
\newblock In Kukla, R., editor, {\em Aesthetics and Cognition in Kant's
  Critical Philosophy}, chapter~7, pages 162 -- 193. Cambridge University
  Press, Cambridge (MA).

\bibitem[Hanslick, 1891]{Hanslick1891}
Hanslick, E. (1891).
\newblock {\em The Beautiful in Music. A Contribution to the Revisal of Musical
  Aesthetics}.
\newblock Novello, London.
\newblock Translated by G. Cohen.

\bibitem[Hertz et~al., 1991]{HertzKroghPalmer91}
Hertz, J., Krogh, A., and Palmer, R.~G. (1991).
\newblock {\em Introduction to the Theory of Neural Computation}, volume~I of
  {\em Lecture Notes of the Santa Fe Institute Studies in the Science of
  Complexity}.
\newblock Perseus Books, Cambridge (MA).

\bibitem[Hochreiter and Schmidhuber, 1997]{HochreiterSchmidhuber97}
Hochreiter, S. and Schmidhuber, J. (1997).
\newblock Long short-term memory.
\newblock {\em Neural Computation}, 9(8):1735 -- 1780.

\bibitem[Huron, 2006]{Huron06}
Huron, D. (2006).
\newblock {\em Sweet Anticipation: Music and the Psychology of Expectation}.
\newblock MIT Press, Cambridge (MA).

\bibitem[Huron, 2019]{Huron19}
Huron, D. (2019).
\newblock Musical aesthetics: Uncertainty and surprise enhance our enjoyment of
  music.
\newblock {\em Current Biology}, 29(23):R1238 -- R1240.

\bibitem[Itti and Baldi, 2005]{IttiBaldi05}
Itti, L. and Baldi, P. (2005).
\newblock Bayesian surprise attracts human attention.
\newblock In Weiss, Y., Sch\"{o}lkopf, B., and Platt, J., editors, {\em
  Advances in Neural Information Processing Systems}, volume~18. MIT Press.

\bibitem[Kant, 2011]{Kant11}
Kant, I. (1764 | 2011).
\newblock {\em Observations on the Feeling of the Beautiful and Sublime}.
\newblock Cambridge Texts in the History of Philosophy. Cambridge University
  Press, Cambridge (MA).

\bibitem[Kant, 1999]{Kant99}
Kant, I. (1787 | 1999).
\newblock {\em Critique of Pure Reason}.
\newblock The Cambridge Edition of The Works Of Immanuel Kant. Cambridge
  University Press, Cambridge, 2nd edition.

\bibitem[Kant, 2015]{Kant15}
Kant, I. (1788 | 2015).
\newblock {\em Critique of Practical Reason}.
\newblock Cambridge Texts in the History of Philosophy. Cambridge University
  Press, Cambridge.

\bibitem[Kant, 1914]{Kant14}
Kant, I. (1790 | 1914).
\newblock {\em Kant's Critique of Judgement}.
\newblock Macmillan, London, 2nd edition.
\newblock Translated by J. H. Bernard.

\bibitem[Kant, 2000]{Kant00}
Kant, I. (2000).
\newblock {\em First Introduction to the Critique of the Power of Judgment}.
\newblock Cambridge University Press, Cambridge (MA).

\bibitem[Kenett et~al., 2023]{KenettHumphriesChatterjee23}
Kenett, Y.~N., Humphries, S., and Chatterjee, A. (2023).
\newblock A thirst for knowledge: Grounding curiosity, creativity, and
  aesthetics in memory and reward neural systems.
\newblock {\em Creativity Research Journal}, 35(3):412 -- 426.

\bibitem[Kim and Sch\"onecker, 2022]{KimSchonecker22}
Kim, H. and Sch\"onecker, D., editors (2022).
\newblock {\em Kant and Artificial Intelligence}.
\newblock De Gruyter, Berlin.

\bibitem[Krumhansl, 1990]{Krumhansl90}
Krumhansl, C.~L. (1990).
\newblock {\em Cognitive Foundations of Musical Pitch}.
\newblock Oxford University Press, New York.

\bibitem[Kukla, 2006]{Kukla06}
Kukla, R., editor (2006).
\newblock {\em Aesthetics and Cognition in Kant's Critical Philosophy}.
\newblock Cambridge University Press, Cambridge (MA).

\bibitem[LeCun et~al., 2015]{CunBengioHinton15}
LeCun, Y., Bengio, Y., and Hinton, G. (2015).
\newblock Deep learning.
\newblock {\em Nature}, 521(7553):436 -- 444.

\bibitem[McCormack and d'Inverno, 2012]{CormackInverno12}
McCormack, J. and d'Inverno, M., editors (2012).
\newblock {\em Computers and Creativity}.
\newblock Springer, Berlin.

\bibitem[Menninghaus et~al., 2019]{MenninghausWagnerEA19}
Menninghaus, W., Wagner, V., Wassiliwizky, E., Schindler, I., Hanich, J.,
  Jacobsen, T., and Koelsch, S. (2019).
\newblock What are aesthetic emotions?
\newblock {\em Psychological Review}, 126(2):171 -- 195.

\bibitem[Meyer, 1956]{Meyer56}
Meyer, L.~B. (1956).
\newblock {\em Emotion and Meaning in Music}.
\newblock University of Chicago Press, Chicago (IL), paperback 1961 edition.

\bibitem[Michaelis, 1795]{Michaelis1795}
Michaelis, C.~F. (1795).
\newblock {\em Ueber den Geist der Tonkunst: Mit R{\"u}cksicht auf Kants Kritik
  der aesthetischen Urtheilskraft. Ein {\"a}sthetischer Versuch}.
\newblock Sch\"afersche Buchhandlung, Leipzig.

\bibitem[Mizraji, 2010]{Mizraji10}
Mizraji, E. (2010).
\newblock {\em En busca de las leyes del pensamiento}.
\newblock Editorial Trilce, Montevideo (Uruguay).

\bibitem[Mizraji, 2023]{Mizraji23}
Mizraji, E. (2023).
\newblock Creating order in the mind: Borges' paradoxical mirror.
\newblock {\em Journal of Genius and Eminence}, 5(2):97 --107.

\bibitem[Mogren, 2016]{Mogren16}
Mogren, O. (2016).
\newblock {C-RNN-GAN: Continuous recurrent neural networks with adversarial
  training}.
\newblock In {\em Proceedings of the Constructive Machine Learning Workshop
  (CML) of the Neural Information Processing Symposium (NIPS)}.

\bibitem[Moore, 1978]{Moore78}
Moore, G.~H. (1978).
\newblock {The origins of Zermelo's axiomatization of set theory}.
\newblock {\em Journal of Philosophical Logic}, 7(1):307 -- 329.

\bibitem[Murray, 2019]{Murray19}
Murray, N. (2019).
\newblock {PFAGAN: An aesthetics-conditional GAN for generating photographic
  fine art}.
\newblock In {\em Proceedings of the IEEE/CVF International Conference on
  Computer Vision Workshop (ICCVW)}, pages 3333 -- 3341.

\bibitem[Ott, 1993]{Ott93}
Ott, E. (1993).
\newblock {\em Chaos in Dynamical Systems}.
\newblock Cambridge University Press, New York.
\newblock Reprinted 1994.

\bibitem[Peitgen and Richter, 1986]{PeitgenRichter86}
Peitgen, H.-O. and Richter, P.~H. (1986).
\newblock {\em The Beauty of Fractals. Images of Complex Dynamical Systems}.
\newblock Springer, Berlin.

\bibitem[Primas, 1990]{Primas90a}
Primas, H. (1990).
\newblock Mathematical and philosophical questions in the theory of open and
  macroscopic quantum systems.
\newblock In Miller, A.~I., editor, {\em Sixty-two Years of Uncertainty:
  Historical, Philosophical and Physics Inquries into the Foundation of Quantum
  Mechanics}, pages 233 -- 257. Plenum Press, New York.

\bibitem[Russell and Norvig, 2010]{RussellNorvig10}
Russell, S. and Norvig, P. (2010).
\newblock {\em Artificial Intelligence: A Modern Approach}.
\newblock Pearson, 3rd edition.

\bibitem[Schmidhuber, 2010]{Schmidhuber10a}
Schmidhuber, J. (2010).
\newblock Formal theory of creativity, fun, and intrinsic motivation (1990 --
  2010).
\newblock {\em IEEE Transactions on Autonomous Mental Development}, 2(3):230 --
  247.

\bibitem[Schmidhuber, 2012]{Schmidhuber12}
Schmidhuber, J. (2012).
\newblock {\em A formal theory of creativity to model the creation of art},
  pages 323 -- 337.
\newblock In \cite{CormackInverno12}.

\bibitem[Schmidhuber, 2015]{Schmidhuber15}
Schmidhuber, J. (2015).
\newblock Deep learning in neural networks: An overview.
\newblock {\em Neural Networks}, 61:85 -- 117.

\bibitem[Silvers, 1996]{Silvers96}
Silvers, R. (1996).
\newblock {\em Photomosaics: Putting Pictures in Their Place}.
\newblock PhD thesis, Massachusetts Institute of Technology (MIT), Cambridge
  (MA).

\bibitem[Smolensky, 1986]{Smolensky86}
Smolensky, P. (1986).
\newblock Information processing in dynamical systems: Foundations of harmony
  theory.
\newblock In Rumelhart, D.~E., McClelland, J.~L., and the PDP Research~Group,
  editors, {\em Parallel Distributed Processing: Explorations in the
  Microstructure of Cognition}, volume~I, chapter~6, pages 194 -- 281. MIT
  Press, Cambridge (MA).

\bibitem[Smolensky, 2006]{Smolensky06}
Smolensky, P. (2006).
\newblock Harmony in linguistic cognition.
\newblock {\em Cognitive Science}, 30:779 -- 801.

\bibitem[Smolensky and Legendre, 2006]{SmolenskyLegendre06a}
Smolensky, P. and Legendre, G. (2006).
\newblock {\em The Harmonic Mind. From Neural Computation to
  Optimality-Theoretic Grammar}, volume 1: Cognitive Architecture.
\newblock MIT Press, Cambridge (MA).

\bibitem[Stangneth, 2019]{Stangneth19}
Stangneth, B. (2019).
\newblock {\em H\"assliches Sehen}.
\newblock Rowohlt, Reimbek.

\bibitem[Sun et~al., 2011]{SunGomezSchmidhuber11}
Sun, Y., Gomez, F., and Schmidhuber, J. (2011).
\newblock {Planning to be surprised: Optimal Bayesian exploration in dynamic
  environments}.
\newblock In {\em Proceedings of the Fourth Conference on Artificial General
  Intelligence (AGI-2011)}.

\bibitem[Tedesco, 2024]{Tedesco24}
Tedesco, S. (2024).
\newblock {Starting from Plessner's ``aesthesiology of the spirit'': Sound and
  normative value of the senses}.
\newblock {\em Mind and Matter}, this issue.

\bibitem[Troll and beim Graben, 1998]{TrollGraben98}
Troll, G. and beim Graben, P. (1998).
\newblock Zipf's law is not a consequence of the central limit theorem.
\newblock {\em Physical Reviews E}, 57(2):1347 -- 1355.

\bibitem[Vear and Poltronieri, 2022]{VearPoltronieri22}
Vear, C. and Poltronieri, F., editors (2022).
\newblock {\em The Language of Creative AI: Practices, Aesthetics and
  Structures}.
\newblock Springer Series on Cultural Computing. Springer, Cham.

\bibitem[Wagner, 1984]{Wagner84}
Wagner, R. (1852|1984).
\newblock {\em Oper und Drama}, volume 8207 of {\em Universalbibliothek}.
\newblock Reclam, Stuttgart.

\bibitem[Williams, 1988]{Williams88}
Williams, R.~J. (1988).
\newblock On the use of backpropagation in associative reinforcement learning.
\newblock In {\em Proceedings of the IEEE International Conference on Neural
  Networks (ICANN)}, volume~1, pages 263 -- 270.

\bibitem[Yang et~al., 2017]{YangChouYang17}
Yang, L.-C., Chou, S.-Y., and Yang, Y.-H. (2017).
\newblock Midinet: A convolutional generative adversarial network for
  symbolic-domain music generation.
\newblock In Hu, X., Cunningham, S.~J., Turnbull, D., and Duan, Z., editors,
  {\em Proceedings of the 18th International Society for Music Information
  Retrieval Conference (ISMIR 2017)}, pages 324 -- 331.
\newblock arXiv:1703.10847 [cs.SD].

\bibitem[Yu et~al., 2021]{YuSrivastavaCanales21}
Yu, Y., Srivastava, A., and Canales, S. (2021).
\newblock {Conditional LSTM-GAN for melody generation from lyrics}.
\newblock {\em ACM Transactions on Multimedia Computation and Communication
  Applications}, 17(1).

\bibitem[Zoeller, 2024]{Zoeller24}
Zoeller, M. (2024).
\newblock How random is random?
\newblock {\em Mind and Matter}, this issue.

\end{thebibliography}


\section*{Appendix}

Here, I present the sources of some of the photomosaic details \Fig{fig:roses}(b): (a) Omega Nebula M17\footnote{
    \url{https://apod.nasa.gov/apod/ap230908.html}
}, (b) Galaxy NGC 7331\footnote{
    \url{https://apod.nasa.gov/apod/ap230914.html}
}, (c) Rose \emph{Komsomolskij Ogonek}\footnote{
    \url{https://commons.wikimedia.org/wiki/File:Rosa_\%27Komsomolskij_Ogonek\%27_Klimenko_1962.jpg}
}, (d) Philosopher Immanuel Kant 1790\footnote{
    \url{https://commons.wikimedia.org/wiki/File:Immanuel_Kant_portrait_c1790.jpg?uselang=de}
}, (e) Author's edited book volume \emph{Lectures in Supercomputational Neuroscience}\footnote{
    \url{https://link.springer.com/book/10.1007/978-3-540-73159-7}
}, (f) Umberto Eco's novel \emph{Il nome della rosa}\footnote{
    \url{https://it.wikipedia.org/wiki/Il_nome_della_rosa#/media/File:9788412451207-scaled-e1636366124330.jpg}
}, (g) Ring Nebula M57\footnote{
    \url{https://en.wikipedia.org/wiki/Messier_object#/media/File:M57_The_Ring_Nebula.JPG}
}, (h) Rose \emph{M\"unsterland}\footnote{
    \url{https://commons.wikimedia.org/wiki/File:Rosa_\%27M\%C3\%BCnsterland\%27_Noack_1986.jpg}
}, (i) Rose \emph{Sacramento}\footnote{
    \url{https://commons.wikimedia.org/wiki/File:Rosa_\%27Sacramento\%27_Anni_Berger_GPG_Langensalza_1981.jpg}
}, (j) Horsehead Nebula IC 434\footnote{
    \url{https://apod.nasa.gov/apod/ap231120.html}
}, (k) Italian semiotician and novelist Umberto Eco\footnote{
    \url{https://de.wikipedia.org/wiki/Umberto_Eco#/media/Datei:Eco,_Umberto-1.jpg}
}, (l) Dan Brown's novel \emph{The Da Vinci Code}\footnote{
    \url{https://danbrown.com/}
}, (m) the same as (c).

\end{document}